\documentclass[twocolumn,showpacs,preprintnumbers,amsmath,amssymb,superscriptaddress]{revtex4}

\usepackage{graphicx}
\usepackage{dcolumn}
\usepackage{bm}

\begin{document}

\preprint{APS/123-QED}

\title{Magneto-transport properties of (Ga, Mn)As close to the metal-insulator transition: Description using Aronov-Altshuler 3D scaling theory}

\author{J. Honolka}
 \email{j.honolka@fkf.mpg.de}
 \affiliation{Max-Planck-Institut f\"ur Festk\"orperforschung, Heisenbergstrasse 1, D-70569 Stuttgart}
 \affiliation{Condensed Matter Physics 114-36, California Institute of Technology, Pasadena, California 91125}
\author{S. Masmanidis}
\affiliation{Condensed Matter Physics 114-36, California Institute
of Technology, Pasadena, California 91125}
\author{H.X. Tang}
\affiliation{Condensed Matter Physics 114-36, California Institute
of Technology, Pasadena, California 91125}
\author{D.D. Awschalom}
\affiliation{Department of Physics, University of California, Santa
Barbara, California 93106}
\author{M.L. Roukes}
\affiliation{Condensed Matter Physics 114-36, California Institute
of Technology, Pasadena, California 91125}

\date{\today}

\begin{abstract}
\noindent The magnitude of the anisotropic magnetoresistance (AMR)
and the longitudinal resistance in compressively strained
(Ga$_{0.95}$, Mn$ _{0.05}$)As epilayers were measured for the first
time down to temperatures as low as 30 mK. Below temperatures of 3K
the conductivity decreases $\propto T^{1/3}$ over two orders of
magnitude in temperature. The conductivity can be well described
within the framework of a 3D scaling theory of Anderson's transition
in the presence of spin scattering in semiconductors. It is shown
that the samples are on the metallic side but very close to the
metal-insulator transition (MIT). At lowest temperatures a decrease
in the AMR effect is observed, which is assigned to changes in the
coupling between the remaining itinerant carriers and the local Mn
5/2-spin moments.
\end{abstract}


\maketitle


Mn-doped diluted III-V semiconductors have been of great interest
since the discovery of a ferromagnetic phase at low
temperatures\cite{Ohno92}. The prospect of using diluted
ferromagnets as spin injectors in spintronic devices have initiated
multiple research efforts to understand their fundamental magnetic
properties and to control important parameters like the magnetic
anisotropy energy or the Curie temperature\cite{Mun93, Haya01,
Pota01}. Though several mechanisms have been proposed to
theoretically describe ferromagnetism in dilute magnetic
semiconductors (DMS) a profound understanding is still elusive. It
is widely accepted though that in the case of GaMnAs holes in the
valence band produced by the Mn-dopants play the key role in
mediating the interaction between spin 5/2 Mn-sites. However, since
ferromagnetism was found in both the metallic and insulating phase
of DMS the correlation between magnetism and transport is expected
to be rather complex. In the metallic regime ferromagnetic coupling
between the local Mn-spins via spin polarization of $p$-type
carriers are often described by Zener type mean-field
models\cite{Dietl01, Koenig03}. Approaching the MIT localization
effects due to disorder become increasingly important, which alter
the magnetic behavior as described by models taking into account
spacial inhomogeneities\cite{Paal91, Kamin02, Berciu01,
Yang03}.\newline Experimentally, the combined study of conductivity
and magneto-transport is a powerful tool to characterize DMS in the
viewpoint of carrier localization and magnetic properties. Most
magneto-transport studies so far were done on the anomalous Hall
effect (AHE) in the high magnetic field regime using classic
Hall-bar geometries with applied fields perpendicular to the
current. A complementary approach to gain information on the
magnetic properties is the study of the anisotropic
magneto-resistance effect. The latter was shown to induce rather
large Hall-resistance jumps up to $80\Omega$ during reorientation of
the magnetization in ferromagnetic GaMnAs epilayers\cite{Tang03}. In
this work we present a study of the conductivity and the AMR
behavior of (Ga$_{0.95}$, Mn$ _{0.05}$)As epilayers for the first
time down to temperatures as low as 30mK. We show that in the lowest
temperature regime the transport can be well described by 3D scaling
theory of Anderson's transition. Our results thus offer an
alternative interpretation to very recent reports on the low
temperature conductivity of metallic GaMn(5$\%$)As epilayers, where
weak localization\cite{Matsukura04} and Kondo models\cite{He05} are
employed. A small reduction of the AMR signal towards lowest
temperatures will be discussed in terms of a localization dependent
ferromagnetic coupling between Mn spins.\newline
\section*{Experimental}
We studied (Ga$_{0.95}$, Mn$ _{0.05}$)As-epilayers of 150nm
thickness grown on a insulating GaAs(001) substrate with a buffer
layer fabricated by means of molecular beam epitaxy at $250^\circ$C.
Because of the compressive strain in the film introduced by the
lattice mismatch with the substrate the material is known to have a
magnetic easy axis in the plane\cite{Ohno99a} with a Curie
temperature of $T_c \approx 45$K in the case of our samples. The
epilayer is patterned into a $100\mu$m wide Hall-bar oriented along
the [110] direction using electron beam lithography. Multiple
voltage probes $100\mu$m apart at both sides of the bar are used to
measure the Hall resistance and the longitudinal resistance per
square (see sketch in Fig.~\ref{loops}). Further experimental
details are given elsewhere~\cite{Tang03}. In-plane magnetic fields
up to $H$ = 2kOe were generated by a superconducting Nb coil. Field
dependent measurements were therefore limited to the temperature
range below the critical temperature of the superconducting coil of
about 5K. The angle $\phi_H$ between the field orientation and the
[110] direction within the plane of the epilayer was set to
15$^\circ$ a indicated in Fig.~\ref{loops}. Special care was taken
to ensure sufficient thermal coupling between the sample and the
cold finger of the dilution refrigerator. To minimize external
rf-heating the entire setup is placed inside a shielded room and,
additionally, leads to the sample are RC-filtered with a cut-off
frequency of a few kHz. For every temperature the amplitude of the
bias-current was chosen to optimize the signal to noise ratio
without heating the sample. At lowest temperatures the currents had
to be reduced to 500pA, while above 1K current amplitudes of 10nA
could be used. Measurements while sweeping the magnetic field have
been obtained in steps of 5Oe with an average ramping speed of 0.4
Oe/s. These low ramping speeds were necessary at lowest temperatures
to avoid heating effects due to Eddy currents.
\section*{Results}
According to AMR theory, both the measured longitudinal resistance
and the Hall resistance depend on the orientation of the
magnetization with respect to the bias current direction. For the
case of a single domain magnetized $100\mu$m $\times$ $100\mu$m
square this can be expressed as
\begin{eqnarray}
R_{\square}  = (\rho_{\parallel} + (\rho_{\parallel} -
\rho_{\perp})\, \text{cos}^{2}\phi)/t
\end{eqnarray}
\begin{eqnarray}
R_{\text H}  = (\rho_{\parallel} - \rho_{\perp})\, \text{sin}\phi\,
\text{cos}\phi/t
\end{eqnarray}
where $R_{\square}$ and $R_{\text H}$ are the longitudinal and the
Hall resistance per square, respectively. $\phi$ is the angle
between the magnetization $M$ and the bias current,
$\rho_{\parallel}(\rho_{\perp}$) the resistivity for currents
parallel (perpendicular) to the magnetization and $t$ the thickness
of the epilayer.
\begin{figure}[b]
\centerline{\includegraphics[width=9cm, clip,angle=0]{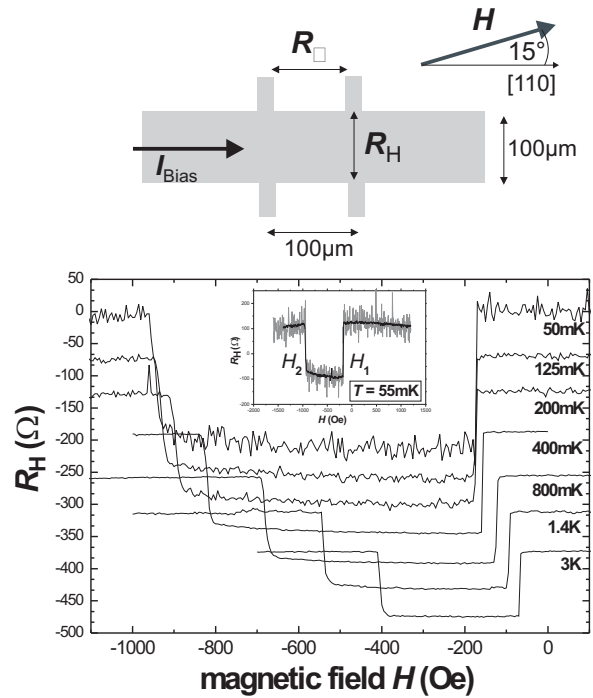}}
\caption{\label{loops}Hall resistance measurements while ramping the
in-plane magnetic field at an average speed of 0.4 Oe/s for
different temperatures. The curves are shifted by a constant offset
with respect to each other. The increased noise at low temperatures
is due to the reduction of the bias current. In the inset two
measurements for bias currents of 100pA (grey) and 1nA (black) at
$T=55$mK are shown for comparison.}
\end{figure}
We want to comment at this point that AMR theories were originally
developed to describe ferromagnetic metals, which will not be
applicable on diluted magnetic semiconductors without modification.
For the case of transition metals with strong exchange splitting of
the the $d$-bands, $(\rho_{\parallel} - \rho_{\perp})$ is a positive
value and is believed to result from anisotropic $sd$-scattering of
minority spin electrons into extended $d$-band states. In (Ga,
Mn)As, however, the spin dependent scattering channels of the
$p$-type hole carriers with e.g. single localized Mn-spins are
different and more complex. This is reflected in the fact that
$(\rho_{\parallel} - \rho_{\perp})$ is experimentally found to be
negative in compressively strained (Ga, Mn)As with Mn concentrations
around 5-8\%, which was recently explained
theoretically\cite{Jung03}.
\newline In Fig.~\ref{loops} the Hall resistance $R_{\text H}(H)$ is shown for
selected temperatures, while the in-plane magnetic field is ramped
from positive to negative values. The two jump behavior of the
curves in Fig.~\ref{loops} can be modeled if one assumes a free
energy density of the form $E = K_{\text u} \text{sin}^2\phi +
(K_{\text 1} / 4) \text{cos}^2 (2\phi) -  M H \text{sin}( \phi
-\phi_H )$ as described elsewhere\cite{Tang03} leading to four local
minima at $\phi_{1,2} = \pm (\pi /4 - \delta)$ and  $\phi_{3,4}  =
\pm (3 \pi/ 4 + \delta )$ for $H = 0$. Here, $K_{\text u}$ and
$K_{\text 1}$ are the in-plane uniaxial and cubic anisotropy
constants and $\delta = \text{sin}^{-1} (K_{\text u} / K_{\text 1}
)$. The two jumps observed in the data in Fig.~\ref{loops} then
reflect the sequential transitions of the magnetization orientation
from [100] ($\phi \sim - 45^{\circ}$) via [010] ($\phi \sim
+45^{\circ}$) to [100] ($\phi \sim +135^{\circ}$).
\begin{figure}[b]
\centerline{\includegraphics[width=9.5cm,
clip,angle=0]{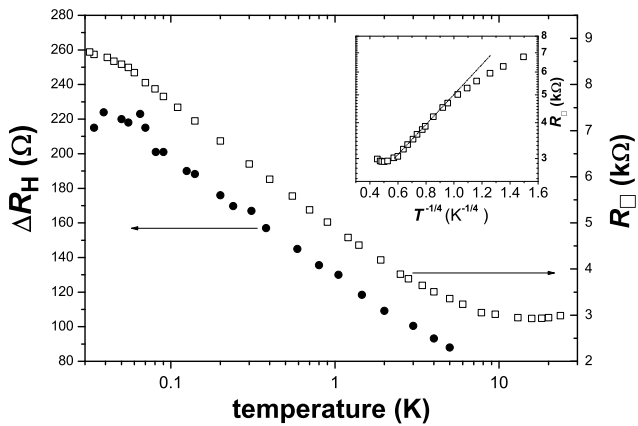}} \caption{\label{resistance}
($\square$) Temperature dependence of the longitudinal resistance
per square measured at a bias current of 1nA and zero magnetic
field.  Prior to the measurement the sample was magnetized at
$H_{\text{S}} = 2$kOe. ($\bullet$) Temperature dependence of the
first Hall resistance jump ($H_1$)
 evaluated at the maximum and minimum values of $R_{\text H}(H)$.}
\end{figure}
During each transition a single 90$^{\circ}$ domain wall travels
through the sample. The magnitude of the observed jumps in $R_{\text
H}$ increases with decreasing temperature and at the same time $H_1$
and $H_2$ are both shifted to larger absolute values. For the
evaluation of the magnitude of the AMR we chose the transition at
$H_1$ since it is more abrupt and therefore better defined compared
to that at $H_2$, which seems to happen more gradually. In order to
exclude Stoner rotation effects, the magnitude of the first jump,
$\Delta R_{\text H}$, was evaluated at the maximum and minimum
values of $R_{\text H}(H)$, which corresponds to well defined values
of $\phi = \pm (\pi/ 4)$. As shown in Fig.~\ref{resistance}, $\Delta
R_{\text H}$ is increasing with decreasing temperature and a maximum
value of 220$\Omega$ is reached at around 40-50mK. Again we want to
emphasize that it was assured that decreasing the magnetic field
ramping speed or bias current did not alter the results of the Hall
resistance measurements even at lowest temperatures (see inset of
Fig.~\ref{loops}).
\newline For comparison the measured longitudinal
resistance $R_{\square}$ at static field conditions $H=0$ is shown
in Fig.~\ref{resistance}. Prior to the resistance measurement the
sample was uniformly magnetized using a saturating field
$H_{\text{S}} = 2$kOe. At about $(16 \pm 2)$K a minimum in the
resistance appears, which is consistent with earlier
measurements\cite{Tang03, Oiwa98}. At lower temperatures the
resistance rises monotoneously and in the range between 1.5K and
50mK seems to follow the temperature dependence of $\Delta R_{\text
H}$. Below 50mK the slope of the resistance starts to flatten
presumably as a consequence of the limited thermal coupling of the
sample.
\begin{figure}[b]
\centerline{\includegraphics[width=8.6cm,
clip,angle=0]{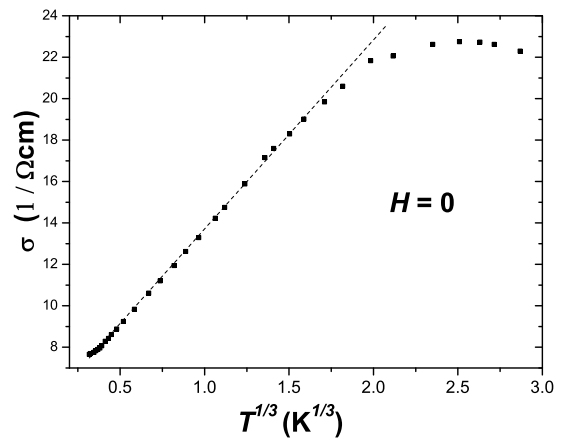}} \caption{\label{conductance}
Conductivity $\sigma=1/R_{\square}t$ plotted versus $T^{1/3}$.
$t=150$nm is the sample thickness. Over two orders of magnitude in
temperature the conductivity is well described by $\sigma=\sigma_0
+\alpha T^{1/3}$ with $\sigma_0=4.5(\Omega\text{cm})^{-1}$ and
$\alpha =9.11 [1/\Omega\text{cm K}^{1/3}]$ (dashed line).}
\end{figure}

\section*{Discussion}
Ferromagnetic (Ga, Mn)As is known to show an increase in resistivity
around the Curie temperature, which is assigned to spin disorder
scattering of holes by spin fluctuation in the phase transition
region\cite{Matsu98, Besch99}. The highest temperature part of the
data in Fig.~\ref{resistance} thus reflects the spin ordering
process below the Curie temperature of about $45$K leading to a
transient decrease in the resistance. At even lower temperatures the
sheet resistance increases again, which is commonly interpreted as
the onset of MIT leading to transport governed by thermally
activated hopping. In this case a behavior $\sim
\text{exp}[-(\frac{T}{T_{0}})^{1/4}]$ is expected assuming variable
range hopping as found by other authors in the temperature range
$7\text{K} > T > 2$K \cite{Oiwa98, Esch97}. The variable range
hopping model, however, fits out data only in a small temperature
window between 1.5K and 8K (see inset of Fig.~\ref{resistance}) and
clearly fails at lowest temperatures. This proves that it is a
different mechanism that leads to a weaker increase of the sheet
resistance with decreasing temperature. From the absence of the
hopping regime we conclude that a larger fraction of carriers stay
delocalized or weakly localized even down to mK-temperatures where
the sheet resistance $R_{\square}$ rises to about 8.5 k$\Omega$,
which is close to the Mott critical value of MIT. In MIT theory the
conductivity $\sigma=1/R_{\square}t$ is the important parameter to
look at, where $t$ is the sample thickness. Fig.~\ref{conductance}
convincingly shows that $\sigma$ exhibits a temperature dependence
$\propto T^{1/3}$ below $T=3$K spanning a range of two orders of
magnitude in temperature. Recently transport measurements above $T =
2$K were interpreted employing weak carrier localization in
3D\cite{Matsukura04} and the Kondo effect\cite{He05}. The authors
predict a temperature dependence $\sigma \propto T^{1/2}$ and
$\propto \text{ln}(T)$, respectively, which is not according to our
observation. Considering that our data covers a much larger
temperature range including for the first time temperatures well
below $1$K we want to suggest an alternative interpretation. A
temperature dependence of the conductivity $\propto T^{1/3}$ was
proposed by 3D scaling theory of Anderson's transition in the
presence of strong spin scattering in semiconductors\cite{Alt83} and
was experimentally found in doped GaAs\cite{Mali88, Romero90,
Capoen93} an Ge\cite{Shlimak96} semiconductors at low temperatures.
According to this theory close to the MIT the Fermi liquid theory
breaks down and $\sigma$ is governed merely by the carrier
correlation length $\xi$ and the carrier interaction length
$L_{\text T} = \sqrt{D\hbar / k_{\text B}T}$. Approaching the MIT
$\xi$ diverges and in the limit $\xi \gg L_T$ the temperature
dependence of the conductivity can be written as~\cite{Mali88}

\begin{equation}
{\sigma=´{2\over 3}{{\text e}^2\over {\hbar \xi}}+{e^2\over
\hbar}({2\over 3\pi}{\partial N\over \partial
\mu}k_{\text{B}}T)^{1/3}}
\label{eq:one}
\end{equation}

The positive zero temperature conductance
$\sigma_0=\sigma(T=0)=4.5(\Omega\text{cm})^{-1}$ deduced in
Fig.~\ref{conductance} is consistent with the values of the order of
$\sim~ 1 (\Omega\text{cm})^{-1}$ found for nGaAs close to the
MIT\cite{Mali88, Romero90, Capoen93}. Positive values indicate that
our sample is still on the metallic side of the MIT as expected. The
slope of the conductivity in Fig.~\ref{conductance} enables an
estimate of the density of states (DOS), $\partial N / \partial
\mu$, at the Fermi level. With $\alpha =9.11 [1/\Omega\text{cm
K}^{1/3}]$ we derive $\partial N / \partial \mu \sim 1 \times
10^{44}[1/\text{Jm}^3]$. For GaMnAs with a higher Mn concentration
of $6.25$\% first principle calculations predict a Mn-4p and As-4p
DOS of about 0.20 and 0.18 [1/eVatom], respectively\cite{Zhao01},
which corresponds to a total DOS of about $1.8\times
10^{46}[1/\text{Jm}^3]$. This is in rather good agreement with our
experimental value considering the lower Mn concentration in our
samples and the fact that the calculations do not take into account
compensation effects due to As${_{\text{Ga}}}$ antisites and Mn
interstitial defects, which are known to reduce the number of holes
by up to 80 \%\cite{Ohno99a}.

\begin{figure}[b]
\centerline{\includegraphics[width=9cm, clip,angle=0]{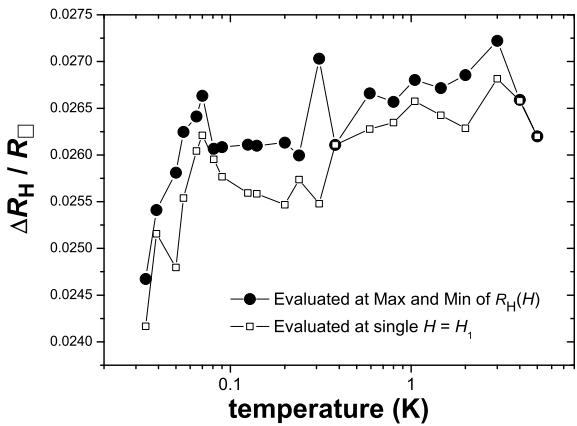}}
\caption{\label{ratio} Temperature dependence of the ratio between
the planar Hall resistance jump $\Delta R_{\text H}$ and the
longitudinal resistance $R_{\square}$. For comparison also an
evaluation of $\Delta R_{\text H}$ at the single field $H_1$ is
shown in open squares. The ratio has a maximum value at around 2K
and decreases by about 6\% towards lowest temperatures.}
\end{figure}

For the case of DMS the regime close to the MIT is currently under
debate and theoretical models have been developed that emphasize the
disorder in the randomly doped and strongly compensated (Ga, Mn)As,
when carriers become localized\cite{Kamin02, Berciu01}. In contrast
to mean-field models these seem to capture details of the measured
magnetic properties like the slight non-mean-field-like concave
shape of the temperature dependence of the magnetization $M$($T$),
which some claim to be present even on the metallic side of the
MIT\cite{DasSarma03b}. It is therefore instructive to look at the
temperature dependence of $\Delta R_{\text H}/R_{\square}(T)$ shown
in Fig.~\ref{ratio}, which gives information about changes in the
relative magnitude of the AMR. While at high temperatures this ratio
increases with decreasing temperature reflecting the increasing spin
alignment of the local Mn-5/2 moments as shown in earlier
works\cite{Tang03}, the situation is different in the range below a
few Kelvin where the longitudinal resistance $R_{\square}$ starts to
rise. The ratio levels off and seems to exhibit a maximum at a
temperature of around $(2\pm1)$K. At even lower temperatures below a
few K, where $\sigma \propto T^{1/3}$ holds, a reduction of the
relative AMR by (6$\pm 2$)\% can be seen. It is conceivable that
this decrease is related to the proximity to the MIT and thus the
onset of carrier localization. As mentioned in the introduction, it
was recently proposed by theoreticians that, due to inhomogeneous
disorder in the randomly doped system, localization of carriers can
affect their magnetic properties. Localization of carriers happens
preferentially in the vicinity of Mn-sites where they are strongly
antiferromagnetically coupled to the adjacent
Mn-spins\cite{DasSarma03b}. These localized holes do not play a role
in the dc transport properties of the DMS such as the AMR but may to
a certain extend alter the coupling between the local moments and
the remaining carriers. A change of the AMR is then expected at
lowest temperatures as hinted by our data.

\section*{Aronov-Altshuler model applied on conductivity data in the literature}
\begin{figure}[b]
\centerline{\includegraphics[width=9cm, clip,angle=0]{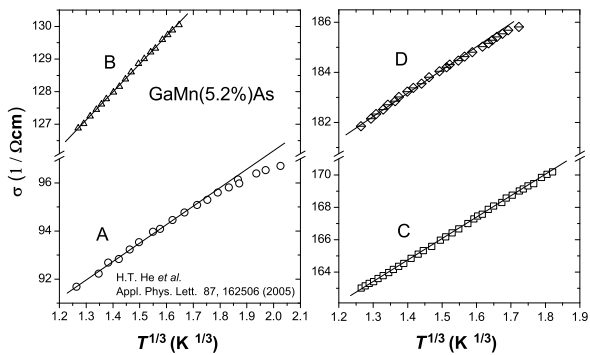}}
\caption{\label{He} Conductivity of GaMn(5.2\%)As measured by H. T.
He {\it et al.} (Appl. Phys. Lett {\bf 87}, 162506 (2005)) for
temperatures below 22K. Samples A, B, C and D: as grown, and
annealed at 160 $^{\circ}$C, 200 $^{\circ}$C and 260$^{\circ}$C,
respectively.}
\end{figure}

\begin{figure}[b]
\centerline{\includegraphics[width=9cm, clip,angle=0]{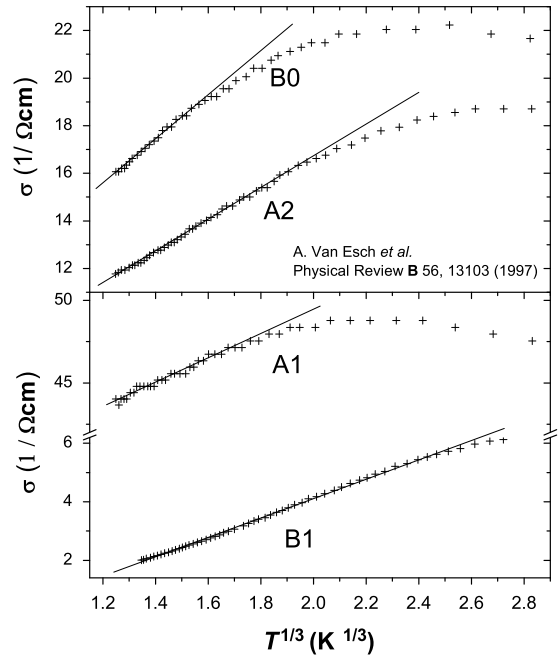}}
\caption{\label{Esch} Conductivity of GaMnAs measured by A. Van Esch
{\it et al.} (Physical Review B {\bf 56}, 13103 (1997)) for
temperatures below 22K. A1 and A2: GaMn(6\%)As annealed at
370$^{\circ}$C and 390$^{\circ}$C, respectively. B0 and B1:
GaMn(7\%)As as grown and annealed at 370$^{\circ}$C, respectively.}
\end{figure}

In this section we want to further support the Aronov-Altshuler
model as an alternative interpretation for the transport machanism
in GaMnAs at low temperatures. The theory will in the following be
applied to existing conductivity data in the literature both on the
insulating and metallic side of the metal insulator transition. In
Fig.~\ref{He} and Fig.~\ref{Esch} the conductivity of GaMnAs thin
films of different Mn concentrations and various annealing
treatments are shown according to He {\it et al.} \cite{He05} and
Van Esch {\it et al.}~\cite{Esch97}. The nomenclature of the samples
corresponds to those of the authors and main sample parameters are
listed in Table~\ref{tab:table1}. Assuming $\sigma=\sigma_0 +\alpha
T^{1/3}$ as proposed by Aronov-Altshuler we can fit the conductivity
rather well below 6K, covering the same temperature range in which
Kondo and variable-range hopping models were proposed~\cite{He05,
Esch97}.
\begin{table}
\caption{\label{tab:table1} Comparison with He {\it et al.} and Van
Esch {\it et al.}}
\begin{ruledtabular}
\begin{tabular}{ccccc}

sample & Mn (\%) & $T_{a}$ ($^{\circ}$C)\footnotemark[1]& $\sigma_0$
$(\Omega\text{cm})^{-1}$ & $\alpha$ $[1/\Omega\text{cm
K}^{1/3}]$ \\
\hline
A & 5.2 & as grown & 92 & 0.3 \\
B & 5.2 & 160 & 116 & 8.6 \\
C & 5.2 & 200 & 146 & 13.1 \\
D & 5.2 & 260 & 169 & 10.3 \\
\hline
A1 & 6.0 & 370 & 35 & 7.3 \\
A2 & 6.0 & 390 & 3.3 & 6.7 \\
B0 & 7.0 & as grown & 4.2 & 9.5 \\
B1 & 7.0 & 370 & -1.7 & 2.7 \\
\end{tabular}
\end{ruledtabular}
\footnotetext[1]{$T_{a}$ denotes the annealing temperature.}
\end{table}
The resulting values for $\sigma_0$ and $\alpha$ are presented in
Table~\ref{tab:table1}. For the case of the metallic GaMn(5.2\%)As
samples A-D (Fig.~\ref{He}), 2h annealing at temperatures up to $T =
260^{\circ}$C increases both $\sigma_0$ and $\alpha$. According to
Eq.~(\ref{eq:one}) this is equivalent to an increase of the DOS at
the Fermi level and a shift away from the MIT towards the metallic
side. This is consistent with the reported reduction of long-range
disorder and defect concentrations upon annealing at 260$^{\circ}$C,
which leads to an increased hole carrier density and a more
effective ferromagnetic coupling between Mn spins.~\cite{Pota01,
Haya01}. The measurements shown in Fig.~\ref{Esch} include samples
with a considerably lower conductivity compared to those of He {\it
et al.}. According to Van Esch {\it et al.} sample B1 is on the
insulator side, B0 and A2 exactly at the MIT and A1 metallic. Again
this is reflected in the values for $\sigma_0$: For B1 the small
negative value of -1.7 $(\Omega\text{cm})^{-1}$ suggests the
insulating phase but close to the MIT. Like for the case of
compensated n-GaAs~\cite{Mali88} it implies that the carrier
interaction length $L_{\text T}$ determines the conductivity just on
the insulating side of the MIT as well. B0 and A2 give small but
positive values indicating metallicity in the proximity of the MIT.
Finally, A1 clearly exhibits a metallic character. Furthermore,
comparing the as grown and annealed samples B0 and B1 one observes
that $\alpha$ and thus the DOS at the Fermi level is reduced upon
annealing at $T = 370^{\circ}$C. It seems that annealing at higher
temperatures has the opposite effect than annealing at $T =
260^{\circ}$C, as shown for samples A-D. This finding is in
agreement with the literature, which reports qualitative changes of
the annealing effects above temperatures $T =
280^{\circ}$C~\cite{Haya01}. In conclusion, this section shows that
despite the limited temperature range of the available data in the
literature the interpretation using the Aronov-Altshuler model lead
to reasonable conclusions about the DOS and metallicity. Changes of
measured hole concentrations with different annealing procedures in
GaMnAs are well reflected.
\section*{Conclusion}
We have studied magnetotransport of epitaxial grown (Ga$_{0.95}$,
Mn$ _{0.05}$)As-epilayers with in-plane magnetic easy axis for the
first time down to temperatures as low as 30mK. Resistance
measurements of our samples show no evidence of strong localization.
Instead a weaker temperature dependence of the conductivity $\propto
T^{1/3}$ is observed, which can be well explained by a 3D scaling
theory of Anderson's transition in the presence of spin scattering
in semiconductors. The results prove that the samples studied are on
the metallic side of but very close to the metal-insulator
transition. Using scaling theory the DOS at the Fermi level was
estimated to be $\partial N / \partial \mu \sim 1 \times
10^{44}[1/\text{Jm}^3]$, which is in good agreement with theoretical
predictions in the literature. The successful application of the
Aronov-Altshuler model on conductivity data by other authors both on
the insulating and metallic side of the metal insulator transition
underlines the potential of this approach. In our measurements of
the temperature dependence of the anisotropic magnetoresistance
effect we observe a reduction of the magnitude by (6$\pm 2$)\% in
the lowest temperature regime, which suggests changes in the
interaction between the Mn-spins and itinerant carriers in the
system. A possible explanation is the observed onset of localization
of carriers in the vicinity of local Mn-spins.

\section*{ACKNOWLEDGMENTS}
The research was supported by the DARPA/SPINS program and the
Deutsche Forschungsgemeinschaft. We want to thank Prof. P. Wigen and
Prof. A. MacDonald for valuable discussions.
\newpage 
\bibliography{Honolka_GPHE_resubmission}

\begin{thebibliography}{26}
\expandafter\ifx\csname natexlab\endcsname\relax\def\natexlab#1{#1}\fi
\expandafter\ifx\csname bibnamefont\endcsname\relax
  \def\bibnamefont#1{#1}\fi
\expandafter\ifx\csname bibfnamefont\endcsname\relax
  \def\bibfnamefont#1{#1}\fi
\expandafter\ifx\csname citenamefont\endcsname\relax
  \def\citenamefont#1{#1}\fi
\expandafter\ifx\csname url\endcsname\relax
  \def\url#1{\texttt{#1}}\fi
\expandafter\ifx\csname urlprefix\endcsname\relax\def\urlprefix{URL }\fi
\providecommand{\bibinfo}[2]{#2}
\providecommand{\eprint}[2][]{\url{#2}}

\bibitem[{\citenamefont{Ohno et~al.}(1992)\citenamefont{Ohno, Munekata, Penney,
  von Moln\'{a}r, and Chang}}]{Ohno92}
\bibinfo{author}{\bibfnamefont{H.}~\bibnamefont{Ohno}},
  \bibinfo{author}{\bibfnamefont{H.}~\bibnamefont{Munekata}},
  \bibinfo{author}{\bibfnamefont{T.}~\bibnamefont{Penney}},
  \bibinfo{author}{\bibfnamefont{S.}~\bibnamefont{von Moln\'{a}r}},
  \bibnamefont{and} \bibinfo{author}{\bibfnamefont{L.}~\bibnamefont{Chang}},
  \bibinfo{journal}{Phys. Rev. Lett.} \textbf{\bibinfo{volume}{68}},
  \bibinfo{pages}{2664} (\bibinfo{year}{1992}).

\bibitem[{\citenamefont{Munekata et~al.}(1993)\citenamefont{Munekata,
  Zaslavski, Fumagalli, and Gambino}}]{Mun93}
\bibinfo{author}{\bibfnamefont{H.}~\bibnamefont{Munekata}},
  \bibinfo{author}{\bibfnamefont{A.}~\bibnamefont{Zaslavski}},
  \bibinfo{author}{\bibfnamefont{P.}~\bibnamefont{Fumagalli}},
  \bibnamefont{and} \bibinfo{author}{\bibfnamefont{R.}~\bibnamefont{Gambino}},
  \bibinfo{journal}{Phys. Rev. Lett.} \textbf{\bibinfo{volume}{63}},
  \bibinfo{pages}{2929} (\bibinfo{year}{1993}).

\bibitem[{\citenamefont{Hayashi et~al.}(2001)\citenamefont{Hayashi, Hashimoto,
  Katsumoto, and Iye}}]{Haya01}
\bibinfo{author}{\bibfnamefont{T.}~\bibnamefont{Hayashi}},
  \bibinfo{author}{\bibfnamefont{Y.}~\bibnamefont{Hashimoto}},
  \bibinfo{author}{\bibfnamefont{S.}~\bibnamefont{Katsumoto}},
  \bibnamefont{and} \bibinfo{author}{\bibfnamefont{Y.}~\bibnamefont{Iye}},
  \bibinfo{journal}{Appl. Phys. Lett.} \textbf{\bibinfo{volume}{78}},
  \bibinfo{pages}{1691} (\bibinfo{year}{2001}).

\bibitem[{\citenamefont{Potashnik et~al.}(2001)\citenamefont{Potashnik, Ku,
  Chun, Berry, Samarth, and Schiffer}}]{Pota01}
\bibinfo{author}{\bibfnamefont{S.}~\bibnamefont{Potashnik}},
  \bibinfo{author}{\bibfnamefont{K.}~\bibnamefont{Ku}},
  \bibinfo{author}{\bibfnamefont{S.}~\bibnamefont{Chun}},
  \bibinfo{author}{\bibfnamefont{J.}~\bibnamefont{Berry}},
  \bibinfo{author}{\bibfnamefont{N.}~\bibnamefont{Samarth}}, \bibnamefont{and}
  \bibinfo{author}{\bibfnamefont{P.}~\bibnamefont{Schiffer}},
  \bibinfo{journal}{Appl. Phys. Lett.} \textbf{\bibinfo{volume}{79}},
  \bibinfo{pages}{1495} (\bibinfo{year}{2001}).

\bibitem[{\citenamefont{Dietl et~al.}(2001)\citenamefont{Dietl, Ohno, and
  Matsukura}}]{Dietl01}
\bibinfo{author}{\bibfnamefont{T.}~\bibnamefont{Dietl}},
  \bibinfo{author}{\bibfnamefont{H.}~\bibnamefont{Ohno}}, \bibnamefont{and}
  \bibinfo{author}{\bibfnamefont{F.}~\bibnamefont{Matsukura}},
  \bibinfo{journal}{Phys. Rev. B} \textbf{\bibinfo{volume}{63}},
  \bibinfo{pages}{195205} (\bibinfo{year}{2001}).

\bibitem[{\citenamefont{K\"onig et~al.}(2001)\citenamefont{K\"onig, Schliemann,
  Jungwirth, and MacDonald}}]{Koenig03}
\bibinfo{author}{\bibfnamefont{J.}~\bibnamefont{K\"onig}},
  \bibinfo{author}{\bibfnamefont{J.}~\bibnamefont{Schliemann}},
  \bibinfo{author}{\bibfnamefont{T.}~\bibnamefont{Jungwirth}},
  \bibnamefont{and}
  \bibinfo{author}{\bibfnamefont{A.}~\bibnamefont{MacDonald}},
  \emph{\bibinfo{title}{in: Electronic Structure and Magnetism of Complex
  Materials}} (\bibinfo{publisher}{edited by D. Singh and D. Papaconstantopolos
  , Springer Verlag}, \bibinfo{address}{Berlin, Germany},
  \bibinfo{year}{2001}).

\bibitem[{\citenamefont{Paalanen and Bhatt}(1991)}]{Paal91}
\bibinfo{author}{\bibfnamefont{M.}~\bibnamefont{Paalanen}} \bibnamefont{and}
  \bibinfo{author}{\bibfnamefont{R.}~\bibnamefont{Bhatt}},
  \bibinfo{journal}{Physica B} \textbf{\bibinfo{volume}{169}},
  \bibinfo{pages}{153} (\bibinfo{year}{1991}).

\bibitem[{\citenamefont{Kaminski and Sarma}(2002)}]{Kamin02}
\bibinfo{author}{\bibfnamefont{A.}~\bibnamefont{Kaminski}} \bibnamefont{and}
  \bibinfo{author}{\bibfnamefont{S.~D.} \bibnamefont{Sarma}},
  \bibinfo{journal}{Phys. Rev. Lett.} \textbf{\bibinfo{volume}{88}},
  \bibinfo{pages}{247202} (\bibinfo{year}{2002}).

\bibitem[{\citenamefont{Berciu and Bhatt}(2001)}]{Berciu01}
\bibinfo{author}{\bibfnamefont{M.}~\bibnamefont{Berciu}} \bibnamefont{and}
  \bibinfo{author}{\bibfnamefont{R.}~\bibnamefont{Bhatt}},
  \bibinfo{journal}{Phys. Rev. Lett.} \textbf{\bibinfo{volume}{87}},
  \bibinfo{pages}{107203} (\bibinfo{year}{2001}).

\bibitem[{\citenamefont{Yang and MacDonald}(2003)}]{Yang03}
\bibinfo{author}{\bibfnamefont{E.}~\bibnamefont{Yang}} \bibnamefont{and}
  \bibinfo{author}{\bibfnamefont{A.}~\bibnamefont{MacDonald}},
  \bibinfo{journal}{Phys. Rev. B} \textbf{\bibinfo{volume}{67}},
  \bibinfo{pages}{155202} (\bibinfo{year}{2003}).

\bibitem[{\citenamefont{Tang et~al.}(2003)\citenamefont{Tang, Kawakami,
  Awschalom, and Roukes}}]{Tang03}
\bibinfo{author}{\bibfnamefont{H.}~\bibnamefont{Tang}},
  \bibinfo{author}{\bibfnamefont{R.}~\bibnamefont{Kawakami}},
  \bibinfo{author}{\bibfnamefont{D.}~\bibnamefont{Awschalom}},
  \bibnamefont{and} \bibinfo{author}{\bibfnamefont{M.}~\bibnamefont{Roukes}},
  \bibinfo{journal}{Phys. Rev. Lett.} \textbf{\bibinfo{volume}{90}},
  \bibinfo{pages}{107201} (\bibinfo{year}{2003}).

\bibitem[{\citenamefont{Matsukura et~al.}(2004)\citenamefont{Matsukura,
  Sawicki, Dietl, Chiba, and Ohno}}]{Matsukura04}
\bibinfo{author}{\bibfnamefont{F.}~\bibnamefont{Matsukura}},
  \bibinfo{author}{\bibfnamefont{M.}~\bibnamefont{Sawicki}},
  \bibinfo{author}{\bibfnamefont{T.}~\bibnamefont{Dietl}},
  \bibinfo{author}{\bibfnamefont{D.}~\bibnamefont{Chiba}}, \bibnamefont{and}
  \bibinfo{author}{\bibfnamefont{H.}~\bibnamefont{Ohno}},
  \bibinfo{journal}{Physica E} \textbf{\bibinfo{volume}{21}},
  \bibinfo{pages}{1032} (\bibinfo{year}{2004}).

\bibitem[{\citenamefont{He et~al.}(2005)\citenamefont{He, Wang, Ge, Wang, Dai,
  and Wang}}]{He05}
\bibinfo{author}{\bibfnamefont{H.}~\bibnamefont{He}},
  \bibinfo{author}{\bibfnamefont{C.}~\bibnamefont{Wang}},
  \bibinfo{author}{\bibfnamefont{W.}~\bibnamefont{Ge}},
  \bibinfo{author}{\bibfnamefont{J.}~\bibnamefont{Wang}},
  \bibinfo{author}{\bibfnamefont{X.}~\bibnamefont{Dai}}, \bibnamefont{and}
  \bibinfo{author}{\bibfnamefont{Y.}~\bibnamefont{Wang}},
  \bibinfo{journal}{Appl. Phys. Lett.} \textbf{\bibinfo{volume}{87}},
  \bibinfo{pages}{162506} (\bibinfo{year}{2005}).

\bibitem[{\citenamefont{Ohno}(1999)}]{Ohno99a}
\bibinfo{author}{\bibfnamefont{H.}~\bibnamefont{Ohno}}, \bibinfo{journal}{J.
  Magn. Magn. Mat.} \textbf{\bibinfo{volume}{200}}, \bibinfo{pages}{110}
  (\bibinfo{year}{1999}).

\bibitem[{\citenamefont{Jungwirth et~al.}(2003)\citenamefont{Jungwirth, Sinova,
  Wang, Edmonds, Campion, Gallagher, Foxon, Niu, and MacDonald}}]{Jung03}
\bibinfo{author}{\bibfnamefont{T.}~\bibnamefont{Jungwirth}},
  \bibinfo{author}{\bibfnamefont{J.}~\bibnamefont{Sinova}},
  \bibinfo{author}{\bibfnamefont{K.}~\bibnamefont{Wang}},
  \bibinfo{author}{\bibfnamefont{K.}~\bibnamefont{Edmonds}},
  \bibinfo{author}{\bibfnamefont{R.~P.} \bibnamefont{Campion}},
  \bibinfo{author}{\bibfnamefont{B.}~\bibnamefont{Gallagher}},
  \bibinfo{author}{\bibfnamefont{C.}~\bibnamefont{Foxon}},
  \bibinfo{author}{\bibfnamefont{Q.}~\bibnamefont{Niu}}, \bibnamefont{and}
  \bibinfo{author}{\bibfnamefont{A.}~\bibnamefont{MacDonald}},
  \bibinfo{journal}{Appl. Phys. Lett.} \textbf{\bibinfo{volume}{83}},
  \bibinfo{pages}{320} (\bibinfo{year}{2003}).

\bibitem[{\citenamefont{Oiwa et~al.}(1998)\citenamefont{Oiwa, Katsumoto, Endo,
  Hirasawa, Iye, Ohno, Matsukura, Shen, and Sugawara}}]{Oiwa98}
\bibinfo{author}{\bibfnamefont{A.}~\bibnamefont{Oiwa}},
  \bibinfo{author}{\bibfnamefont{S.}~\bibnamefont{Katsumoto}},
  \bibinfo{author}{\bibfnamefont{A.}~\bibnamefont{Endo}},
  \bibinfo{author}{\bibfnamefont{M.}~\bibnamefont{Hirasawa}},
  \bibinfo{author}{\bibfnamefont{Y.}~\bibnamefont{Iye}},
  \bibinfo{author}{\bibfnamefont{H.}~\bibnamefont{Ohno}},
  \bibinfo{author}{\bibfnamefont{F.}~\bibnamefont{Matsukura}},
  \bibinfo{author}{\bibfnamefont{A.}~\bibnamefont{Shen}}, \bibnamefont{and}
  \bibinfo{author}{\bibfnamefont{Y.}~\bibnamefont{Sugawara}},
  \bibinfo{journal}{Phys. Stat. Sol.} \textbf{\bibinfo{volume}{205}},
  \bibinfo{pages}{167} (\bibinfo{year}{1998}).

\bibitem[{\citenamefont{Matsukura et~al.}(1998)\citenamefont{Matsukura, Ohno,
  Shen, and Sugawara}}]{Matsu98}
\bibinfo{author}{\bibfnamefont{F.}~\bibnamefont{Matsukura}},
  \bibinfo{author}{\bibfnamefont{H.}~\bibnamefont{Ohno}},
  \bibinfo{author}{\bibfnamefont{A.}~\bibnamefont{Shen}}, \bibnamefont{and}
  \bibinfo{author}{\bibfnamefont{Y.}~\bibnamefont{Sugawara}},
  \bibinfo{journal}{Phys. Rev. B} \textbf{\bibinfo{volume}{57}},
  \bibinfo{pages}{R2037} (\bibinfo{year}{1998}).

\bibitem[{\citenamefont{Beschoten et~al.}(1999)\citenamefont{Beschoten,
  Crowell, Malajovich, Awschalom, Matsukura, Shen, and Ohno}}]{Besch99}
\bibinfo{author}{\bibfnamefont{B.}~\bibnamefont{Beschoten}},
  \bibinfo{author}{\bibfnamefont{P.}~\bibnamefont{Crowell}},
  \bibinfo{author}{\bibfnamefont{I.}~\bibnamefont{Malajovich}},
  \bibinfo{author}{\bibfnamefont{D.}~\bibnamefont{Awschalom}},
  \bibinfo{author}{\bibfnamefont{F.}~\bibnamefont{Matsukura}},
  \bibinfo{author}{\bibfnamefont{A.}~\bibnamefont{Shen}}, \bibnamefont{and}
  \bibinfo{author}{\bibfnamefont{H.}~\bibnamefont{Ohno}},
  \bibinfo{journal}{Phys. Rev. Lett.} \textbf{\bibinfo{volume}{83}},
  \bibinfo{pages}{3073} (\bibinfo{year}{1999}).

\bibitem[{\citenamefont{Esch et~al.}(1997)\citenamefont{Esch, Bockstal, Boeck,
  Verbanck, van Steenbergen, Wellmann, Grietens, Bogaerts, Herlach, and
  Borghs}}]{Esch97}
\bibinfo{author}{\bibfnamefont{A.~V.} \bibnamefont{Esch}},
  \bibinfo{author}{\bibfnamefont{L.~V.} \bibnamefont{Bockstal}},
  \bibinfo{author}{\bibfnamefont{J.~D.} \bibnamefont{Boeck}},
  \bibinfo{author}{\bibfnamefont{G.}~\bibnamefont{Verbanck}},
  \bibinfo{author}{\bibfnamefont{A.}~\bibnamefont{van Steenbergen}},
  \bibinfo{author}{\bibfnamefont{P.}~\bibnamefont{Wellmann}},
  \bibinfo{author}{\bibfnamefont{B.}~\bibnamefont{Grietens}},
  \bibinfo{author}{\bibfnamefont{R.}~\bibnamefont{Bogaerts}},
  \bibinfo{author}{\bibfnamefont{F.}~\bibnamefont{Herlach}}, \bibnamefont{and}
  \bibinfo{author}{\bibfnamefont{G.}~\bibnamefont{Borghs}},
  \bibinfo{journal}{Phys. Rev. B} \textbf{\bibinfo{volume}{56}},
  \bibinfo{pages}{13103} (\bibinfo{year}{1997}).

\bibitem[{\citenamefont{Altshuler and Aronov}(1983)}]{Alt83}
\bibinfo{author}{\bibfnamefont{B.}~\bibnamefont{Altshuler}} \bibnamefont{and}
  \bibinfo{author}{\bibfnamefont{A.}~\bibnamefont{Aronov}},
  \bibinfo{journal}{JETP} \textbf{\bibinfo{volume}{37}}, \bibinfo{pages}{410}
  (\bibinfo{year}{1983}).

\bibitem[{\citenamefont{Maliepaard et~al.}(1988)\citenamefont{Maliepaard,
  Pepper, Newbury, and Hill}}]{Mali88}
\bibinfo{author}{\bibfnamefont{M.}~\bibnamefont{Maliepaard}},
  \bibinfo{author}{\bibfnamefont{M.}~\bibnamefont{Pepper}},
  \bibinfo{author}{\bibfnamefont{R.}~\bibnamefont{Newbury}}, \bibnamefont{and}
  \bibinfo{author}{\bibfnamefont{G.}~\bibnamefont{Hill}},
  \bibinfo{journal}{Phys. Rev. Lett.} \textbf{\bibinfo{volume}{61}},
  \bibinfo{pages}{369} (\bibinfo{year}{1988}).

\bibitem[{\citenamefont{Romero et~al.}(1990)\citenamefont{Romero, Liu, Drew,
  and Ploog}}]{Romero90}
\bibinfo{author}{\bibfnamefont{D.}~\bibnamefont{Romero}},
  \bibinfo{author}{\bibfnamefont{S.}~\bibnamefont{Liu}},
  \bibinfo{author}{\bibfnamefont{H.}~\bibnamefont{Drew}}, \bibnamefont{and}
  \bibinfo{author}{\bibfnamefont{K.}~\bibnamefont{Ploog}},
  \bibinfo{journal}{Phys. Rev. B} \textbf{\bibinfo{volume}{42}},
  \bibinfo{pages}{3179} (\bibinfo{year}{1990}).

\bibitem[{\citenamefont{Capoen et~al.}(1993)\citenamefont{Capoen, Biskupsi, and
  Briggs}}]{Capoen93}
\bibinfo{author}{\bibfnamefont{B.}~\bibnamefont{Capoen}},
  \bibinfo{author}{\bibfnamefont{G.}~\bibnamefont{Biskupsi}}, \bibnamefont{and}
  \bibinfo{author}{\bibfnamefont{A.}~\bibnamefont{Briggs}},
  \bibinfo{journal}{J. Phys.: Condens. Matter} \textbf{\bibinfo{volume}{5}},
  \bibinfo{pages}{2545} (\bibinfo{year}{1993}).

\bibitem[{\citenamefont{Shlimak et~al.}(1996)\citenamefont{Shlimak, Kaveh,
  Ussyshkin, Ginodman, and Resnick}}]{Shlimak96}
\bibinfo{author}{\bibfnamefont{I.}~\bibnamefont{Shlimak}},
  \bibinfo{author}{\bibfnamefont{M.}~\bibnamefont{Kaveh}},
  \bibinfo{author}{\bibfnamefont{R.}~\bibnamefont{Ussyshkin}},
  \bibinfo{author}{\bibfnamefont{V.}~\bibnamefont{Ginodman}}, \bibnamefont{and}
  \bibinfo{author}{\bibfnamefont{L.}~\bibnamefont{Resnick}},
  \bibinfo{journal}{Phys. Rev. Lett.} \textbf{\bibinfo{volume}{77}},
  \bibinfo{pages}{1103} (\bibinfo{year}{1996}).

\bibitem[{\citenamefont{Zhao et~al.}(2001)\citenamefont{Zhao, Geng, Park, and
  Freeman}}]{Zhao01}
\bibinfo{author}{\bibfnamefont{Y.-J.} \bibnamefont{Zhao}},
  \bibinfo{author}{\bibfnamefont{W.}~\bibnamefont{Geng}},
  \bibinfo{author}{\bibfnamefont{K.}~\bibnamefont{Park}}, \bibnamefont{and}
  \bibinfo{author}{\bibfnamefont{A.}~\bibnamefont{Freeman}},
  \bibinfo{journal}{Phys. Rev. B} \textbf{\bibinfo{volume}{64}},
  \bibinfo{pages}{35207} (\bibinfo{year}{2001}).

\bibitem[{\citenamefont{Sarma et~al.}(2003)\citenamefont{Sarma, Hwang, and
  Kaminski}}]{DasSarma03b}
\bibinfo{author}{\bibfnamefont{S.~D.} \bibnamefont{Sarma}},
  \bibinfo{author}{\bibfnamefont{E.}~\bibnamefont{Hwang}}, \bibnamefont{and}
  \bibinfo{author}{\bibfnamefont{A.}~\bibnamefont{Kaminski}},
  \bibinfo{journal}{Phys. Rev. B} \textbf{\bibinfo{volume}{67}},
  \bibinfo{pages}{155201} (\bibinfo{year}{2003}).

\end{thebibliography}

\end{document}